\documentclass{article}

\usepackage{arxiv}

\usepackage[utf8]{inputenc} 
\usepackage[T1]{fontenc}    
\usepackage{hyperref}       
\usepackage{url}            
\usepackage{booktabs}       
\usepackage{amsfonts}       
\usepackage{nicefrac}       
\usepackage{microtype}      
\usepackage{lipsum}
\usepackage{graphicx}
\usepackage{multirow}
\graphicspath{ {./images/} }

\title{Cybersecurity Challenges in the Offshore Oil and Gas Industry: An Industrial Cyber-Physical Systems (ICPS) Perspective}

\author{
 Abubakar Sadiq Mohammed \\
  School of Computer Science and Informatics\\
  Cardiff University, UK\\
  \texttt{mohammedas@cardiff.ac.uk} \\
   \And
 Philipp Reinecke \\
  School of Computer Science and Informatics\\
  Cardiff University, UK\\
  \texttt{reineckep@cardiff.ac.uk} \\
  \And
 Pete Burnap \\
  School of Computer Science and Informatics\\
  Cardiff University, UK\\
  \texttt{burnapp@cardiff.ac.uk} \\
  \And
 Omer Rana \\
  School of Computer Science and Informatics\\
  Cardiff University, UK\\
  \texttt{ranaof@cardiff.ac.uk} \\
  \And
 Eirini Anthi \\
  School of Computer Science and Informatics\\
  Cardiff University, UK\\
  \texttt{anthies@cardiff.ac.uk} \\
}

\begin{document}
\maketitle
\begin{abstract}
The offshore oil and gas industry has recently been going through a digitalisation drive, with use of `smart' equipment using technologies like the Industrial Internet of Things (IIoT) and Industrial Cyber-Physical Systems (ICPS). There has also been a corresponding increase in cyber attacks targeted at oil and gas companies. Oil production offshore is usually in remote locations, requiring remote access and control. This is achieved by integrating ICPS, Supervisory, Control and Data Acquisition (SCADA) systems, and  IIoT technologies. A successful cyber attack against an oil and gas offshore asset could have a devastating impact on the environment, marine ecosystem and safety of personnel. Any disruption to the world's supply of oil and gas (O\&G) can also have an effect on oil prices and in turn, the global economy. This makes it important to secure the industry against cyber threats. We describe the potential cyberattack surface within the oil and gas industry, discussing emerging trends in the offshore sub-sector, and provide a timeline of known cyberattacks. We also present a case study of a subsea control system architecture typically used in offshore oil and gas operations and highlight potential vulnerabilities affecting the components of the system. This study is the first to provide a detailed analysis on the attack vectors in a subsea control system and is crucial to understanding key vulnerabilities, primarily to implement efficient mitigation methods that safeguard the safety of personnel and the environment when using such systems. 
\end{abstract}


\section{Introduction}
Despite worldwide initiatives to implement green energy sources, the global demand for crude oil is expected to remain high for decades to come~\cite{dnv2017oilandgas}, \cite{wanasinghe2020internet}. This makes it ever more important to protect the industry from increasing cyber threats. While no business is immune to cyber security attacks, critical industries such as oil and gas (O\&G) are increasingly more vulnerable, with many hackers now targeting (directly or indirectly) the operational domain \cite{avanzini2019cybersecurity}. This is because Operational Technology (OT) and Industrial Control Systems (ICS) in the past had been physically isolated from outside networks \cite{bundi2020effects} and based on proprietary hardware, software, and communications protocols \cite{drias2015analysis}. This ensured that the only risk they were exposed to was within the scope of their operations locally \cite{stouffer2011guide} as many ICS components were in physically secured areas and the components were not connected to Information Technology (IT) networks or systems. However, due to the need to make faster and better business decisions, equipment are now smarter and are being integrated with multiple industrial technologies in IT in line with Industry 4.0 \cite{stergiopoulos2020cyber}, sometimes referred to as "Oil and Gas 4.0" \cite{lu2019oil}, \cite{progoulakis2021perspectives}. This integration supports new IT capabilities, but it provides significantly less isolation for ICS from the outside world than predecessor systems \cite{stouffer2011guide}. As a result, the attack surface has widened with attacks growing in frequency over the last few years.

\begin{figure}[ht]
    \centering
    \includegraphics[width=0.75\textwidth]{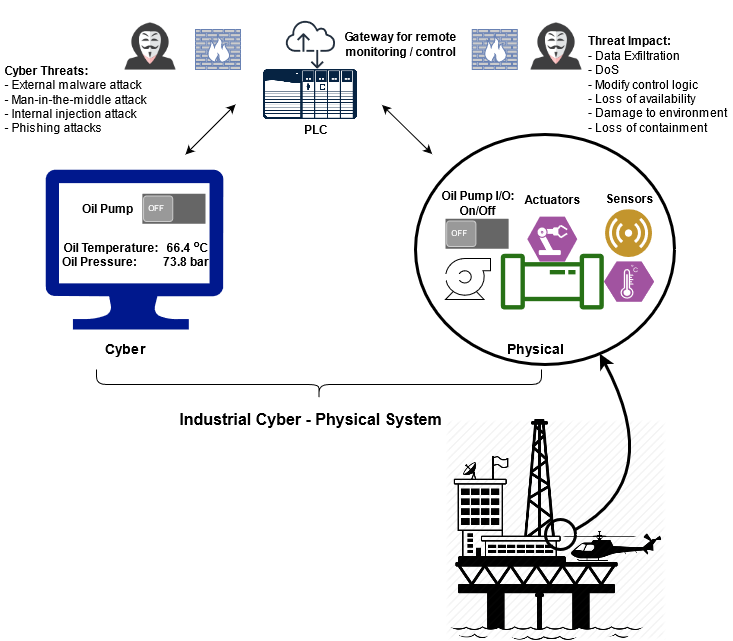}
    \caption{Example of a Cyber-Physical System in oil and gas highlighting common vulnerabilities}
    \label{fig:cyber_physical_system}
\end{figure}

The integration of OT and IT has been aided by the rapid development of embedded systems, sensors, and networks, which in turn, has given rise to Cyber-Physical Systems (CPS) capable of advanced computing and networking technologies in a unified way \cite{kayan2021cybersecurity}. Industrial Cyber-Physical System (ICPS) refers to CPS that is specifically designed for industrial applications \cite{kayan2021cybersecurity}. This has opened the door to significant efficiency gains in the oil and gas industry\cite{settemsdal2019go} and is particularly the  case  in  the  offshore  sector,  where  there  is  a  pressing  need  to  reduce  costs  and  maximize  equipment availability \cite{settemsdal2019go}. While it allows engineers to monitor and control assets remotely \cite{stergiopoulos2020cyber}, \cite{alcaraz2013critical}, \cite{stellios2018survey}, this exposes ICS communication protocols to cyberspace vulnerabilities, like data exfiltration and malware injection attacks,  which could cause significant losses to a company and potentially compromise process safety; endangering lives of personnel including damage to the environment. The migration to IT has also led to the standardization of new SCADA communication protocols such as Modbus-TCP Distributed Network Protocol (DNP3), IEC-60870–5-104 and the Inter-Control Center Protocol (lCCP, IEC60870–6) \cite{alcaraz2013critical}. The first three were designed for automation and control, and the last was designed to connect SCADA systems \cite{alcaraz2013critical}. Figure \ref{fig:cyber_physical_system} shows an example of some typical components that make up an ICPS setup in an offshore O\&G platform and highlights common vulnerabilities.

Being heavy industrial processes, O\&G production and processing facilities rely heavily on ICPS \cite{nygaard2020dragonstone}. Control equipment such as Programmable Logic Controllers (PLCs) and Distributed Control Systems (DCS) are widely used, along with Human Machine Interfaces (HMIs) and Remote Terminal Units (RTUs) \cite{nygaard2020dragonstone}. Using Industrial Internet of Things (IIoT) technology, the interconnection of these intelligent industrial devices with control and management platforms, collectively improve the operational efficiency and productivity of industrial systems \cite{xu2018survey}. One of the more common use cases of ICPS in the O\&G industry that depend on these control equipment is Asset Performance Management (APM) as it is a data-driven approach to asset management \cite{settemsdal2019go}. APM solutions are often linked to Computerized Maintenance Management Systems (CMMS) which also drives their deployment on-premise \cite{settemsdal2019go}.As the move toward minimally manned facilities continues, having remote visibility into operations becomes increasingly important \cite{settemsdal2019go}. Currently, the need to transfer production data to information systems, which also includes the needs for remote maintenance \cite{nygaard2020dragonstone} has helped in broadening the attack surface across the O\&G industry.

Due to their remote location in deep waters and the need for real time monitoring and control, offshore O\&G assets, have potentially a wider attack surface compared to other sub-sectors of the industry, which makes them attractive to threat actors. This is critical because offshore production accounts for a significant proportion (about 30\% \cite{offshoreproduction2015}) of global O\&G production. There also seems to be a passive shift in focus towards offshore production in some oil producing countries. In Nigeria, for example, some International Oil Companies (IOCs) are divesting their onshore producing assets to focus more on deep offshore production \cite{nigeriadivestment2021} \cite{hellenicdivestment2021}. In America, Equinor, which has a large portfolio of offshore assets in the US Gulf of Mexico, has agreed to divest its onshore assets in the Bakken Field \cite{equinordivestment2021}. These trends indicate that the offshore O\&G sub-sector is likely to retain or increase its share of global oil and gas production.

Successful cyber-attacks threaten the competitiveness of the global O\&G industry, and the cost of future breaches will be much higher, whether to corporate assets, public infrastructure and safety, or the broader economy through energy prices \cite{ciepiela2016digitization}. Breaches can lead to lost production, raised health, safety, and environmental risk, costly damages claims, breach of insurance conditions, negative reputational impacts, and loss of licence to operate. Therefore, cybersecurity needs to be a consideration throughout the life-cycle of any project, especially across digital transition activity \cite{dickinson2019guest}.

The reported percentages of acknowledged cyber attacks indicate the high threat for offshore oil and gas assets \cite{progoulakis2021perspectives}. A cyber-attack on an O\&G OT environment can have serious results beyond just financial losses including environmental damage, loss of human lives \cite{ciepiela2016digitization}, data and information theft, direct manipulation of machinery \cite{stergiopoulos2020cyber}, changes to inclination of entire oil rigs \cite{stergiopoulos2020cyber}, \cite{space2016tilting} or pressurisation of pipelines \cite{stergiopoulos2020cyber}, \cite{lee2014mediareport}, \cite{robertson2014mysterious}.

\subsection{O\&G vs Other Critical Infrastructure Industries}
\begin{figure}
    \centering
    \includegraphics[width=0.6\textwidth]{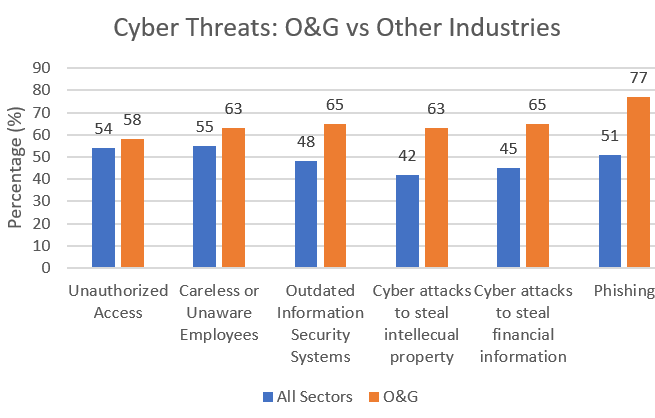}
    \caption{Cyber threats faced by O\&G sector as compared to other industrial sectors. Source: EY Global Information Security Survey 2016-17 \cite{ciepiela2016digitization}}
    \label{fig:chart_comparison}
\end{figure}

In 2017, EY carried out a 2016-17 Global Information Security Survey shown in Figure \ref{fig:chart_comparison} where selected companies were asked which threats and vulnerabilities have most increased their risk exposure over the last 12 months. In every single metric recorded, the O\&G companies had a higher cyber-attack incidence occurring compared to other critical infrastructure industries. More specifically, in the United States, the Department of Homeland Security responded to more than 350 cyber attack incidents at US energy companies between 2011 and 2015 and identified nearly 900 security vulnerabilities within those energy companies - a figure that was higher than any other industry \cite{nygaard2020dragonstone} \cite{eaton2018hacked}.

Another study that examined the state of cybersecurity in the United States O\&G industry was carried out by The Ponemon Institute \cite{ponemon2017stateof} in 2017, where 377 individuals who were responsible for securing or overseeing cyber risk in the OT environment were surveyed. It was discovered that only 41\% continuously monitor all infrastructure to prioritize threats and attacks. An average of 46\% of all cyber attacks in the OT environment go undetected, suggesting the need for investments in technologies that detect cyber threats to O\&G operations \cite{ponemon2017stateof}.

The vulnerability of this sector was evident in May 2021, when one of the largest pipelines in the US which carries refined gasoline and jet fuel from Texas up the East Coast to New York, was forced to shut down its 5,500 miles of pipelines for six days due to a cybersecurity attack \cite{uspipeline2021}. Reports indicate that this was a ransomware attack that targeted the IT system, yet its repercussions were felt in OT operations as headline news reported panic, social disruption, and a crippling lack of fuel delivery \cite{reeder2021cybersecurity}.

\subsection{Related Work}

\begin{table}[]
\begin{tabular}{|l|p{6cm}|c|c|}
\hline
Author/Reference & Summary & Multiple Sectors & O\&G Specific \\ \hline \hline
Alcaraz et al. \cite{alcaraz2013critical} & Presented generic architectural   components of critical infrastructure components and their vulnerabilities & \checkmark &  \\ \hline
Kim et al. \cite{kim2012cyber} & Surveyed CPS research in   multiple domains including hybrid systems, security, and real-time computing   and outlined potential for CPS in several applications & \checkmark &  \\ \hline
Krotofil et al. \cite{krotofil2013industrial} & Presented a survey on ICS   security research including security controls to mitigate vulnerability of   common ICS protocols. & \checkmark &  \\ \hline
Mo et al. \cite{mo2011cyber} & Their survey highlighted   information security and system-theory-based security approach to securing   cyber-physical systems & \checkmark &  \\ \hline
Stergiopoulos et al. \cite{stergiopoulos2020cyber} & Presented an attack taxonomy and   catalogue of cyber attacks on O\&G assets &  & \checkmark \\ \hline
McLaughlin et al. \cite{mclaughlin2016cybersecurity} & Presented an overview of ICS   security assessment including the key principles of ICS operations. & \checkmark &  \\ \hline
Sadeghi et al. \cite{sadeghi2015security} & Authors examine security and   privacy issues relating to IIOT with proposed mitigations & \checkmark &  \\ \hline
Stellios et al. \cite{stellios2018survey} & Authors assessed the IIoT threat   landscape by analysing representative attacks against IoT in a risk-like   approach & \checkmark &  \\ \hline
Khan et al. \cite{khan2017reliable} & Proposed IoT architecture   specifically for the O\&G industry to aid functional and business   requirements. The alternate architecture was based on three modules   applicable to the upstream, midstream, and downstream sectors &  & \checkmark \\ \hline
Sayegh et al. \cite{sayegh2013internal} & Authors presented a testbed used   to detect vulnerabilties in SCADA protocols to internal attacks. & \checkmark &  \\ \hline
Nazir et al. \cite{nazir2017assessing} & Authors survey tools and   techniques to discover SCADA system vulnerabilities common to CPS deployed in   numerous sectors & \checkmark &  \\ \hline
Bhamare et   al. \cite{bhamare2020cybersecurity} & Explored major publications from   industry and academia and addressed applicability of machine learning   techniques for ICS cybersecurity & \checkmark &  \\ \hline
Miller et al. \cite{miller2012survey} & Authors analysed cyber security   incidents on critical infrastructure and SCADA systems and developed a   taxonomy to classify future SCADA security incidents & \checkmark &  \\ \hline
Giraldo et al. \cite{giraldo2017security} & Authors lay out a classification   in CPS domains, security level implementation, and computational strategies   from a survey of numerous surveys & \checkmark &  \\ \hline
\end{tabular}
\caption{Summary of Related Work}
\label{tab:related_work}
\end{table}

A number of publications have outlined potential attacks on SCADA systems conventionally used in the O\&G industry. These are summarised in Table \ref{tab:related_work}. Very few publications survey cybersecurity topics specifically for the O\&G sector \cite{stergiopoulos2020cyber}, although some papers~\cite{mcbride2016overload}, \cite{dragos2019global}, \cite{lobo2018upstream} survey cybersecurity incidents related to the O\&G industry \cite{stergiopoulos2020cyber}. Stergiopoulos et al. \cite{stergiopoulos2020cyber} were the first to develop a vulnerability taxonomy for ICPS specifically for the O\&G sector. The concepts presented in the study are broadly applicable to the 3 sub-sectors in O\&G: upstream, downstream and midstream. We extend their study by analysing the key components in a subsea control system -- which are usually designed to different standards from onshore platforms due to the extreme conditions that exist in deep waters, and we highlight vulnerabilities of the system to cyber attacks. The research to date has focused generally on ICS security, which is broadly applicable to most sectors with only a few analysing real cyber security incidents that have taken place in the O\&G sector. Studies on O\&G ICPS security have lacked domain knowledge of a complete end-to-end process system, while highlighting vulnerabilities. This is important because showing vulnerabilities of specific existing engineering designs used in the field could lead to more resilient systems. From a literature review perspective, it is evident that the subject of cyber security for O\&G assets is not widely studied\cite{progoulakis2021perspectives}. Reports also indicate that the industry's cyber maturity is relatively low, and O\&G boards show very little understanding of cyber issues \cite{lamba2018protecting}, \cite{stergiopoulos2020cyber}. As a result of the increase in cyber attacks on O\&G ICPS and the lack of an understanding of the associated attack vectors in such systems, it is critical to investigate this further in order to ensure better protection of vulnerable assets. 

\subsection{Motivation}
Out of 42 recorded cyber security incidents \cite{stergiopoulos2020cyber} affecting the O\&G industry in the past decade, the upstream sector had the highest number of incidents. This gives an indication of a higher vulnerability in this sector compared to other O\&G sub-sectors. Moreover, because the upstream sector is the first stage of 3 highly inter-connected sectors of the O\&G industry (which will be described briefly in Section \ref{oil_and_gas_overview}), any disruption will likely cascade down the value chain and have an impact on the other sectors. For these reasons, our paper will be focused on the upstream O\&G sector. This work aims to answer key questions such as what unique challenges make the industry more vulnerable to attacks compared to others, and why the available datasets for cyber security research are largely not representative of the O\&G industry processes. To the best of our knowledge, there has been no survey carried out specifically for the offshore environment of the O\&G industry, identifying inherent vulnerabilities in an end-to-end subsea system. 
This paper aims to describe the oil and gas production process and its vulnerabilities, present a timeline of documented cyber attacks on O\&G upstream assets, analyse a subsea control system architecture highlighting the vulnerabilities of the system to cyber attacks, and discuss limitations in available datasets for security research on OT infrastructure. 

Our main contibutions are: (i) A timeline of documented cyber attacks on O\&G upstream assets; (ii) A case study of a subsea control system architecture highlighting the vulnerabilities of the system to cyber attacks; (iii) Mitigation strategies against cyber threats to subsea control systems. The remainder of this paper is structured as follows: Section 2 gives an overview of the upstream sector of the O\&G industry and describes the oil and gas production process. Section 3 discusses common vulnerabilities, attack vectors in the sector, and a case study of a subsea control system architecture is presented. In Section 4, we explain challenges and analyse why upstream O\&G assets are difficult to secure, while Section 5 looks at the state of securing O\&G assets including datasets available for security research. Section 6 is the conclusion.

\section{Overview of the Oil and Gas Industry} \label{oil_and_gas_overview}
\begin{figure}[ht]
    \centering
    \includegraphics[width=0.7\textwidth]{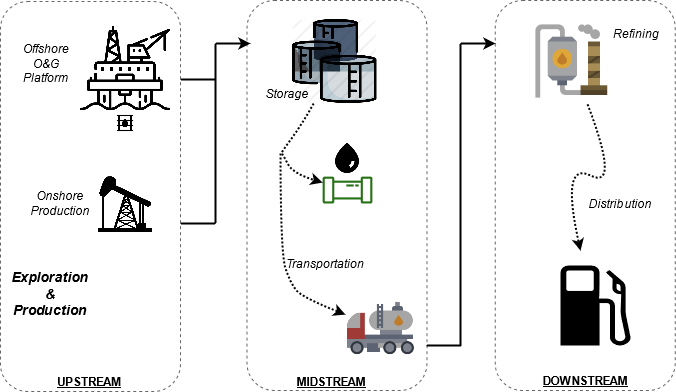}
    \caption{Oil and Gas Value Chain}
    \label{fig:oil_and_gas_value_chain}
\end{figure}

The O\&G industry comprises of three sub-sectors - upstream, downstream, and midstream infrastructures \cite{stergiopoulos2020cyber}. These sectors are quite diverse in their roles within the value chain. The \textbf{\textit{upstream}} sector deals with exploration, drilling, and production \cite{stergiopoulos2020cyber}  - basically all activities involving the search for oil/gas, the recovery process, and production from reservoirs hundreds of feet underneath the Earth’s surface at very high pressures and temperatures. It comprises of offshore and onshore operations. The \textbf{\textit{downstream}} sector focuses on distributing assets to consumers \cite{stergiopoulos2020cyber} and handles the refining of the natural gas or crude oil produced and its storage facilities (oil refineries, Liquefied Natural Gas plants, gas stations, petrochemical plants, etc) while the \textbf{\textit{midstream}} connects the upstream activities to the downstream activities \cite{stergiopoulos2020cyber} (transportation – pipelines, crude oil tankers, trucks; and some marketing activities). These three sectors are interconnected and interact through a complex web of activities which are streamlined to ensure a timely and safe delivery of petroleum products to the end consumers. These sectors are highlighted in Figure \ref{fig:oil_and_gas_value_chain}. In the next sub-sections we will describe the life cycle of an O\&G upstream asset and the oil and gas production process.

\subsection{Life Cycle of the Upstream Oil and Gas Industry}
Upstream activities include exploration, drilling, and production and are typically referred to as E\&P (Exploration \& production). More specifically, the upstream life cycle is split into five phases which cover the 'cradle to grave' activities ranging from how the hydrocarbons are discovered to reservoir depletion, and decommissioning (returning the environment to its pre-E\&P state). The activities that take place in each phase and their average timelines are summarised in Table \ref{tab:lifecycle}.

\begin{center}
\begin{table}[h]
\begin{tabular}{ | l | c | p{9cm} |}
\hline
 Phase & Timing & Activities\\
 \hline \hline
 1. Exploration & 1-5 years & Exploration for potentially viable oil and gas sources through geological surveys. Often  no potentially  viable oil and gas sources are  discovered  and  operations  are terminated \cite{darko2014short}. \\
 \hline
 2. Appraisal & 4-5 years & Sites identified as potentially containing viable oil/gas sources are examined in more  detail \cite{darko2014short}. \\
 \hline
 3. Development & 4-10 years & Limited  infrastructure  and  site  development  will already  be  in  place  as  part  of  the  exploratory  and  initial  drilling  phase,  but during the field development phase activity will dramatically increase and first oil/gas will be produced towards the end of this phase \cite{darko2014short}. \\
 \hline
 4. Production & 20-50 years & Oil/gas  reserves  are  being  extracted  and  transported  for  processing  and distribution \cite{darko2014short}. \\
 \hline
 5. Decommissioning & 2-10 years & Usually during decommissioning, the platform is completely removed and the seafloor returned to its unobstructed pre-lease condition \cite{bull2019worldwide}. Once  it  is  no  longer  cost-effective  to  extract  remaining  reserves,  the  site  is decommissioned  and  the  operating  companies  are  typically  responsible  for returning the site to as close to original state as possible \cite{darko2014short}. \\
 \hline
\end{tabular}
\caption{Upstream life cycle describing activities carried out during each phase}
\label{tab:lifecycle}
\end{table}
\end{center}

\subsection{Upstream O\&G Production and Processing}
\begin{figure}[ht]
    \centering
    \includegraphics[width=0.75\textwidth]{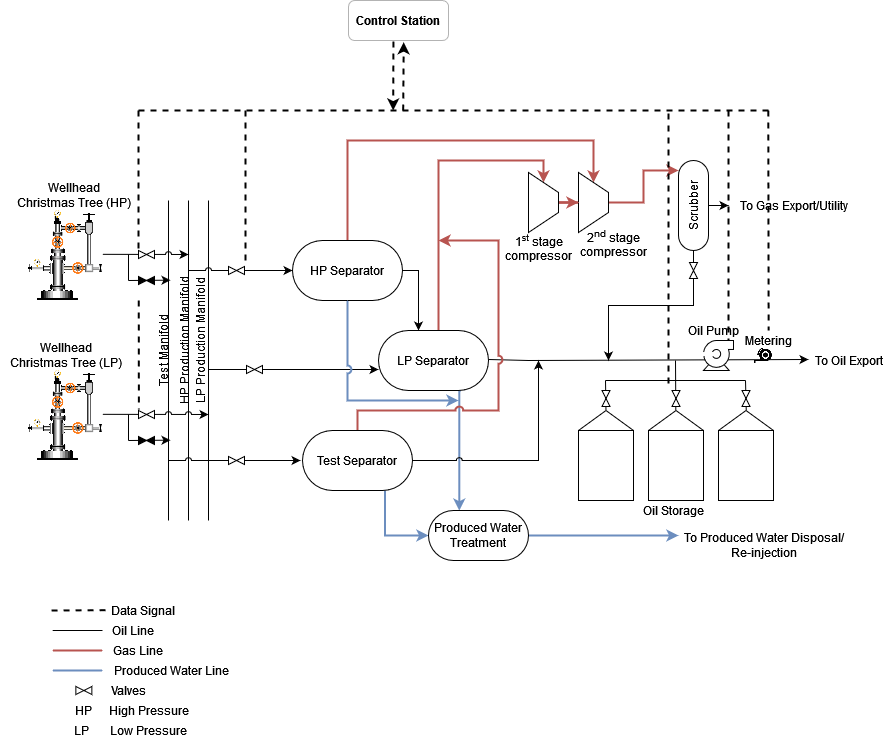}
    \caption{Upstream oil and gas production process}
    \label{fig:upstream_OG_production}
\end{figure}

Oil and gas production is the process of extracting reservoir fluids (hydrocarbons) from beneath the earth's surface -- typically at high pressures and temperatures, and separating the mixture of oil, gas, and water at the surface. The main activities are gathering (from wellheads to separators), separations (separate oil, gas, and water), gas compression (prepare for storage and transport), temporary oil storage, waste water disposal, and metering (calculation of quantity before export) \cite{polyakov2015sap}. From the wellheads, reservoir fluids are fed into production and test manifolds. Next stage is the separation process, where horizontal gravity separators are usually used \cite{devold2006oil} in most facilities. The fluids are separated based on their densities (water is heavier than oil while gas is the lightest). In the separator, the pressure is often reduced in several stages - from high pressure to low pressure - to allow controlled separation of volatile components \cite{devold2006oil}. The gas is dehydrated, compressed and used to power the plant in most cases while the rest is exported. The oil is also processed and stored in settling tanks ready for export while the produced water could be re-injected into the reservoir for pressure maintenance or disposed of safely. There are a number of variations to this process depending on the crude oil composition and the required end products, but this is the typical baseline setup for most oil and gas production facilities. This process is illustrated in Figure \ref{fig:upstream_OG_production}.

\subsection{Offshore Operations}
Offshore O\&G operations are a subset of upstream operations. It is common for offshore O\&G operators to have a service territory that spans a large geographic area \cite{nygaard2020dragonstone}. A large O\&G company operating offshore, for instance, generates, transmits, and stores petabytes of sensitive and competitive field data; and operates and shares thousands of drilling and production control systems spread across geographies, fields, vendors, service providers, and partners \cite{mittal2017}. Most of the field data transmitted and stored are collected by sensors that are part of an industrial control system. ICPS sits at the heart of remote operations which enables the satellite platforms to be fully automated. For this reason, central operations centres may be constructed to control system flow and monitor system conditions \cite{nygaard2020dragonstone}, which is made possible by utilising ICPS for collection of data and control of critical system processes. A large offshore oilfield development project would typically have several types of platforms to effectively extract and export oil and gas resources from the deep oceans. These structures would be distributed around the field(s) (several kilometers apart) as satellite platforms. The reservoir fluids extracted would be transported via pipelines to a central processing facility (CPF) where they are processed, stored, then offloaded to export tankers. The extraction of crude oil from offshore facilities is made possible by subsea control systems, and in recent times subsea production systems. These are highly advanced equipment designed to operate under extreme pressures and temperatures found in deep waters.

\textbf{Drilling Campaigns:} Throughout the life of a field, there will be several drilling campaigns carried out. During the exploration phase, the aim of drilling is to find commercial quantities of hydrocarbons. In the appraisal phase, the aim of drilling is to confirm how large the reservoir is and its characteristics. Development phase drilling is more precise as this is where the initial production wells will be drilled. During production, however, there will still be some drilling campaigns - referred to as in-fill drilling. This is to improve the efficiency of depleting the reservoir by adding more wells during the life of the field. This is why there are usually drilling rigs moving between fields to execute drilling campaigns all over the world. When a drilling rig arrives on site, it can be attached to a host platform for shared resources. This is usually taken into account when designing offshore structures. Drilling operations are usually carried out by oil service companies. They are different from the company who owns and operates the asset. A common use case for ICPS during drilling operations is to enable leak detection, in which a remote multi-sensing technology \cite{chen2017offshore} could be used. This helps in identifying potential leaks and aids quick response to limit the release of harmful hydrocarbons into the environment.

\subsection{Emerging Trends - Remote Offshore O\&G Production Operations}
While most offshore platforms are still currently manned facilities, there seems to be a trend indicating a shift towards operating oil rigs completely remotely from land. Several recent studies and innovations supporting unmanned O\&G production have also indicated this shift in thinking \cite{tan2020transforming} \cite{krishna2020alternative} \cite{mahmoud2020safe} \cite{okpala2020enabling} \cite{vinnem2021assessment}. Some examples of such studies are the DNV GL’s unmanned floating LNG (Liquified Natural Gas) concept, Solitude and  Aker Solutions' conceptual idea for an unmanned FPSO with annual maintenance campaigns \cite{frostad2020unmanned}. Equipment is modularized and monitored from shore with much of the routine maintenance and fault correction carried out by self-programming autonomous inspection and maintenance units \cite{dnv2017oilandgas}. With these developments, it is safe to conclude that the ability to operate an unmanned platform as part of a portfolio  of offshore assets allows materially reduced OPEX \cite{oga2017analysisofukcs}. This is a huge factor influencing O\&G companies operating offshore to invest in this technology, which also has the potential to increase the attack vectors.

\section{Common Vulnerabilities and Attack Vectors in the Upstream O\&G Industry}
In Section \ref{oil_and_gas_overview}, the basic process flow in oil and gas production was described. There are inherent vulnerabilities that an attacker could exploit in the system. In this section, we will discuss the types of attacks that can compromise the system and present a case study of subsea control, communications, and its common vulnerabilities. Process monitoring and remote control are two common activities that utilise ICPS and automation to optimise operations. These are generally applicable to:

\begin{enumerate}

    \item \textbf{Monitoring (Sensors):} Temperature, pressure, chemical composition, leak detection, etc.
    
    \item \textbf{Remote Control:} Valves/actuators, pumps, hydraulic and pneumatic control systems, Safety Instrumented Systems (SIS), Emergency Shutdown Systems (ESD), Fire \& Gas Systems (F\&G), High Integrity Pressure Protection System (HIPPS), etc.

\end{enumerate}

The following sub-section will examine the attacks that could compromise any of these listed operations.

\subsection{Types of Attacks}
 
\begin{figure}[ht]
    \centering
    \includegraphics[width=0.8\textwidth]{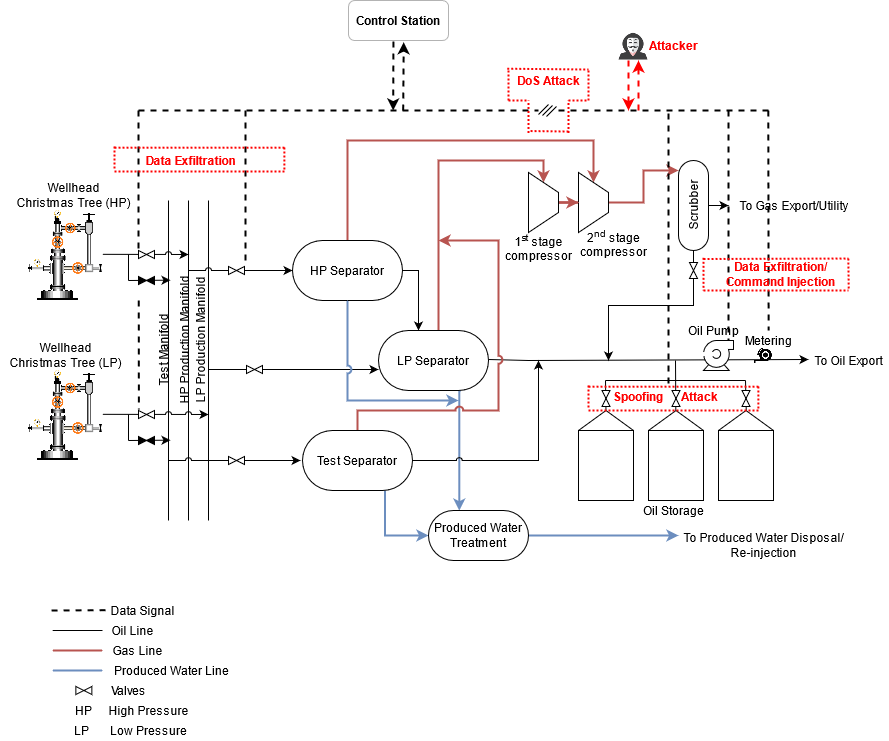}
    \caption{Upstream oil and gas production process showing potential cyber attacks}
    \label{fig:upstream_process_vuln}
\end{figure}


\begin{itemize}
    
    \item \textbf{Denial of Service (DoS):} One of the main safety features for process control are Emergency ShutDown systems (ESD) which are used to prevent unsafe operating conditions. In a DoS attack, an attacker could take advantage of vulnerabilities in data signal like insecure communication protocols not requiring authentication, and flood the network with random commands which effectively renders the ESD incapable of responding to unsafe process control requests. If an attacker, for example, were to carry out a DoS attack on the ESD of an unmanned offshore oil facility, a major catastrophic event could happen if there was a pressure build-up in the crude oil export lines. This kind of attack ensures that the onshore control centre loses its ability to shut down critical process to avert danger.
    
    \item \textbf{Oil Tank Level Spoofing Attack:} Processed oil that has been treated and separated from gas and water is stored in settling tanks ready for export. These tanks are fitted with level control sensors that transmit information to prevent tank overfills. The main goal of this attack is to falsify sensor readings indicating that the tank level is lower than it actually is, which could lead to explosions due to a tank overfill as oil is a highly volatile product.
    
    \item \textbf{Wellhead Production Data Exfiltration:} By discretely deploying malicious software such as trojans on compromised workstations in the control station, an attacker could be privy to sensitive information like wellhead production data. There are various ways a threat actor could harvest sensitive company data they are all stealthy techniques. An example is with the use of Domain Generation Algorithms (DGA) in establishing communications between bots and their Command-and-Control (C\&C) servers. Accessing metering data at custody transfer points also avails an attacker with sensitive information. This could allow threat actors to study hydrocarbon export volumes over time and arm them with enough information to prepare stealthy spoofing attacks that could cause loss of revenue to the company. Some companies have had their data discretely exfiltrated for years before it was found out.
    
     \item \textbf{Command Injection:} PLCs control numerous operations in the oil production process described earlier in Figure \ref{fig:upstream_OG_production}. An example is the oil export system which comprises of export pumps, flow computers, flow meters, and actuators. If an attacker were to compromise an engineering workstation in the control centre, they could alter commands to cause the pump or actuators to perform inappropriately. In addition, PLCs are programmed to control the process to perform within safe operational parameters like maximum allowable pressure and flowrate. These set point limits, if tampered could lead to unsafe operational states. Oil and gas, being volatile products, need very little instability to ignite and cause explosions.
     
     \item \textbf{Data Tampering:} Processed data could be tampered with by an attacker. An  attacker could obfuscate the details of a wider attack by altering operation log and system control-related data \cite{zhang2019multilayer}, which would deceive defenders carrying out a post-attack forensic analysis. Data historians in offshore control stations that store operation log files could be targeted by this attack.
     
     \item \textbf{Choke Size Replay Attack:} In this attack, the signed packets sent over the network could be captured and resent multiple times to the destination \cite{sengupta2020comprehensive}. An example of a dangerous application is if an attacker were to intercept commands sent to increase or decrease the choke size of a well (to increase or decrease crude oil production rates). They could replay these commands to increase the choke size, masking as a legitimate command, which could damage the reservoir permanently.
    
\end{itemize}
Table \ref{tab:attack_motive_impact} summarises some of these potential attacks on upstream O\&G processes showing attacks, attacker motives, vulnerable components, and potential consequences including impact of attack. These vulnerable points in the oil production process are also highlighted in Figure \ref{fig:upstream_process_vuln}.

\begin{table}[]
\begin{tabular}{|p{2cm}|p{2.1cm}|p{2cm}|p{2cm}|p{3cm}|p{3cm}|}
\hline
O\&G   Process & Attacker Motive & Potential Attack & Component & Consequence & Impact \\ \hline \hline
Oil tank   storage & Service disruption & Spoofing & Level sensors & Tank overfill, loss of   containment & Explosion, loss of life,   environmental damage \\ \hline
Hydrocarbon   separation & Revenue loss & Data tampering & Pressure or Temperature sensors & Incomplete separation of gas   from oil & low quality product, loss of   revenue \\ \hline
Oil   delivery, export, piping & Service disruption & Command injection & PLC, pumps, actuators & Operations outside allowable   limits & Potential damage to asset and   environment, potential loss of lives \\ \hline
Emergency   Shutdown & Damage to asset & DoS & Safety Instrumented   System, PLC, and actuators & Operations outside allowable   limits & Potential damage to asset and   environment, potential loss of lives \\ \hline
Custody   Transfer/metering & Revenue loss, theft of   operational information & Data exfiltration & flow computers, meters,   pressure/temperature sensors & Incorrect calculation of   hydrocarbon volumes, sensitive operational data leakage & Loss of revenue, reputational   damage \\ \hline
\end{tabular}
\caption{Attacks on some O\&G upstream processes showing attacker motives and impact}
\label{tab:attack_motive_impact}
\end{table}

\begin{figure}[ht]
    \centering
    \includegraphics[width=12cm]{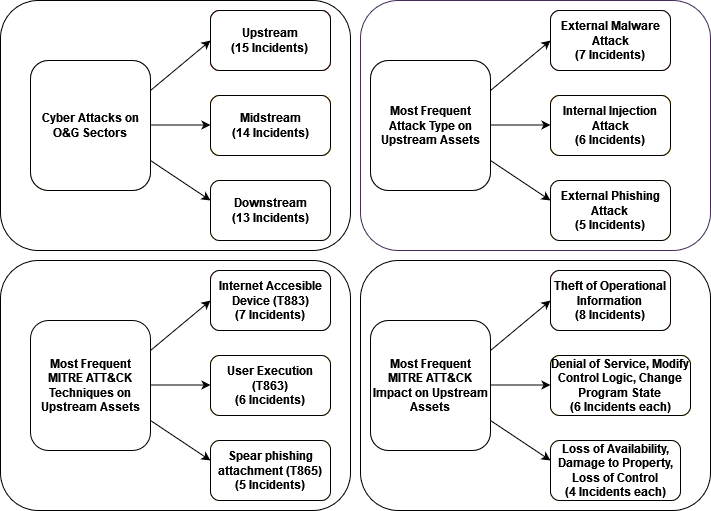}
    \caption{Analysis of cyber attacks on upstream assets \cite{stergiopoulos2020cyber}}
    \label{fig:upstream_MITRE_attacks}
\end{figure}

Based on the recorded security incidents \cite{stergiopoulos2020cyber} affecting upstream O\&G assets, Figure \ref{fig:upstream_MITRE_attacks} shows a pattern that indicates that these vulnerabilities are already being exploited, and that threat actors have this capability. The most frequent impact from these attacks were theft of operational information (8 incidents), Denial of Service (6 incidents), modification of control logic (6 incidents), and change of program state (6 incidents). The impact of these types of attacks on a subsea control system are investigated in sub-section~\ref{sec:subseacasestudy}.

\subsection{Subsea Control and Remote Monitoring - A Case Study}
\label{sec:subseacasestudy}

\begin{figure}[ht]
    \centering
    \includegraphics[width=\textwidth]{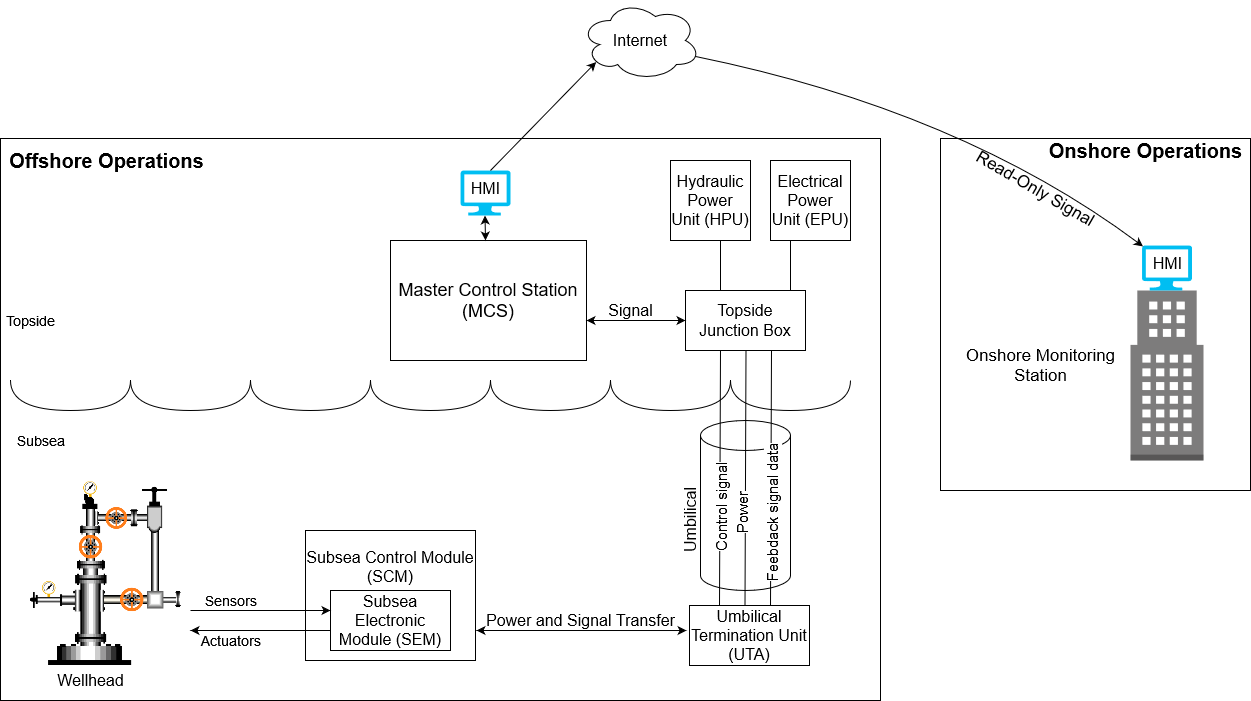}
    \caption{Example of an offshore subsea production monitoring system}
    \label{fig:susbea_production_system}
\end{figure}

One of the critical processes in offshore operations is the subsea control system. Located hundreds of kilometres under deep waters, this system is essentially responsible for real time monitoring of production parameters to prevent unsafe conditions. We have focused on an offshore system because, as discussed earlier, records indicate that the upstream sub-sector is more vulnerable to attacks. Furthermore, an attack on a physical process in an offshore (possibly unmanned) asset will take a much longer time to respond to, compared to an onshore asset, given that it is located hundreds of kilometres in the oceans - which increases the likelihood of devastating impact. An example of a typical setup is shown in Figure \ref{fig:susbea_production_system} where production is monitored via HMIs at the Master Control Station (MCS) and remote workstations that could be located further away in onshore offices. The functions of these components are described in Table \ref{tab:subsea_components}. A subsea control system comprises of one or more of the following components \cite{mudrak2016subsea}:

\begin{itemize}
    \item a wellhead with connected casing strings;
    \item a subsea christmas tree comprising pressure and flow control valves;
    \item a production control and monitoring system for remote monitoring and control of various subsea equipment, possibly multi-phase flow meters;
    \item a chemical injection system;
    \item an umbilical cable with electrical power and signal cables, as well as conduits for hydraulic control fluid and various chemicals to be injected into the produced fluid streams.
\end{itemize}

\begin{table}[]
\begin{tabular}{|l|p{10cm}|}
\hline
Component & Function \\ \hline
Wellhead Christmas Tree & Combines with wellhead to constitute the pressure barrier between reservoir and environment and allow for control of well through various valves and sensors \\ \hline
Subsea Electronic Module (SEM) & Collects   sensor data from wellhead interfaces \\ \hline
Subsea Control Module (SCM) & Houses   the SEM and control valve module \\ \hline
Umbilical Cable & Houses a   collection of hydraulic, data (fibre optic), power cables \\ \hline
Master Control Station (MCS) & Main   field control station where HMIs and servers are located for logging and   processing real time system data \\ \hline
Topside Junction Box & Combines   all electric and hydraulic power generated topside and transmits to subsea   network (umbilical termination unit) \\ \hline
Hydraulic Power Unit (HPU) & Power   source for hydraulics to move valve actuators \\ \hline
Electrical Power Unit (EPU) & Power   source for electrical components \\ \hline
\end{tabular}
\caption{Functions of Some Components of a Subsea Control System}
\label{tab:subsea_components}
\end{table}

The components in the subsea architecture can be split into the following layers \cite{stergiopoulos2020cyber}:

\begin{enumerate}

\item \textbf{Hardware:} Sensors, actuators, RTUs, PLCs, server equipment (racks, CPUs), routers, access control hardware (smart cards, RFID, etc), and valves.

\item \textbf{Firmware:} Operating systems, data and instructions for controlling the hardware.

\item \textbf{Software:} HMIs, Application Programming Interface (APIs), proprietary software packages, and applications.

\item \textbf{Network:} Communications protocols, modems/routers, firewalls

\item \textbf{Process:} Designed ICS business logic, control systems configuration

\end{enumerate}

\subsubsection{Attack Vectors of Subsea Control Systems}

\begin{enumerate}

    \item \textbf{Interception of Commands and Sensor Readings:} The initial stages of an attack requires gathering information on the system and operating parameters. This could be executed with a Man-in-The-Middle (MiTM) attack, where the connection between source and destination ports is intercepted, creating two new channels of communication: one connection between the source device and attacker, and another one between the attacker and the destination device \cite{anthi2017secure}. This attack could compromise the software layer of the subsea architecture. Assuming an attacker managed to compromise a workstation within the MCS, they would gain access to the HMI and sensitive information like pressure and temperature values, production flowrates, maximum allowable working pressure (MAWP), and valve fail-safe positions. Additionally, the attacker is able to act as a proxy and therefore read, insert, and modify data in the intercepted communication \cite{anthi2017secure}. Earlier analysis has shown documented cyber attacks involving theft of operational information as indicated in figure~\ref{fig:upstream_MITRE_attacks}. Adding authentication and encryption of data helps to defend against this kind of threat. Park and Kang \cite{park2016mutual} proposed a solution to MiTM attacks by authentication inter-device communication where each sensor is involved in the generation and distribution of session keys \cite{sengupta2020comprehensive}.
    
    \item \textbf{Injecting Falsified Sensor Data:} The goal of this attack is to compromise the integrity of the sensor readings. This is a spoofing attack which is a variant of the MiTM attack where the attacker modifies data between two communicating devices. The firmware and hardware layers of the subsea architecture are susceptible to this kind of attack. An attacker intercepting communication between the SCM and the MCS conveying sensor readings could modify these values even before the gateways (serial-to-ethernet converters) convert the data to ethernet packets \cite{weiss2019sensors}. Another example is where an attacker, using the compromised workstation, manages to modify control logic of the SCM altering upper or lower limits of set pressure points which could cause a well blowout. The oil spill in the Gulf of Mexico in 2010 has shown how devastating the impact of a subsea well blowout can be to the environment \cite{white2012impact} and safety of personnel. Figure \ref{fig:upstream_MITRE_attacks} shows six incidents of documented cyber attacks each involving modification of control logic and changing program state which indicates that threat actors have this capability. A number of studies have suggested mitigation against this type of attack by using physics-based methods \cite{giraldo2018survey} which consider the effects of the attack on the controlled physical process and look for deviations from expected physical sensor measurements \cite{azzam2021grounds}. Azzam et al. \cite{azzam2021grounds} proposed an Early Warning System (EWS) that, on its own, is not capable of detecting injection of false sensor readings, but can generate early warnings in ICPS based on preliminary indicators. They applied their framework to Linear Time-Invariant (LTI) systems and adapted existing reachability analysis tools to compute a suspicion metric. This could prove useful if integrated with other intrusion detection capabilities to thwart stealthy malicious attempts.
    
    \item \textbf{Denial of Service (DoS) Attack:} One of the most common attacks for cyber adversaries to conduct is the DoS attack \cite{taylor2017security}. System availability is of utmost importance in a subsea control system architecture and the attacker can flood the communicating device with requests to jam the communication channels and prevent legitimate requests \cite{taylor2017security}. DoS can compromise the network and hardware layers of the subsea control system and render engineers in the MCS incapable of sending emergency shutdown commands to shut-in wells discovered to be operating in unsafe conditions. Figure \ref{fig:upstream_MITRE_attacks} shows documented cyber attack cases of DoS and loss of availability in six and four separate incidences affecting upstream O\&G facilities respectively. DoS can have very serious impact by disabling critical equipment in a subsea control system architecture. Sicari et al. \cite{sicari2018reato} proposed a defence mechanism against different types of DoS attacks named REATO. They examined a cross-domain adn flexible middleware, named NetwOrked Smart object (NOS) and tailored REATO to it.

\end{enumerate}

The most common communication protocols being used in this setup are EthernetIP and Modbus. Availability of these systems is key and the communication of both control signals and sensor monitoring data are often not encrypted and not signed for data integrity \cite{hacquebord2019drilling}. With the growth in offshore E\&P activities due to rising number of mature (depleted) onshore oilfields in recent years \cite{mordor2021subseamarket} subsea production is set to dominate a significant market share in the industry. The major vendors in the subsea control equipment market are Subsea 7, Technip FMC, Akastor ASA, Baker Hughes, and National-Oilwell Vargo Inc \cite{mordor2021subseamarket} while for DCS we have ABB, Emerson, Honeywell, Rockwell Automation, Schneider Electric, and Siemens \cite{research2019dcs} dominating the market share. In isolation, these equipment are robust and are safe for operations. However, in a bid to increase their market share, these key vendors controlling the market share are designing products with more and more integration with corporate IT systems which introduces more attack vectors with an increased risk of zero day attacks.

\subsection{History of Cyber-Attacks on Upstream O\&G Assets}
\begin{figure}[ht]
    \centering
    \includegraphics[width=0.85\textwidth]{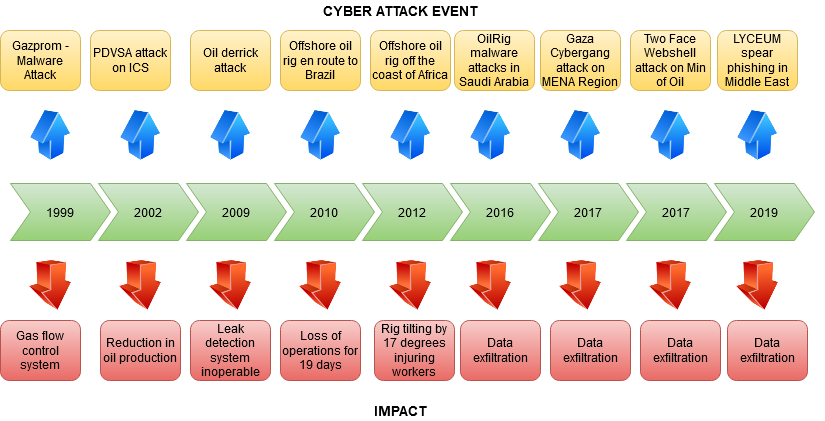}
    \caption{Timeline of Cyber Attacks on Upstream Oil and Gas Facilities \cite{stergiopoulos2020cyber}}
    \label{fig:timeline_upstream_cyberattacks}
\end{figure}

A number of cases have been reported where upstream systems were directly or indirectly compromised by malicious insiders or
malware, causing a number of adverse effects on operations and machinery \cite{lobo2018upstream}, \cite{stergiopoulos2020cyber}, \cite{kravets2009hackers}. Stergiopoulos et al. \cite{stergiopoulos2020cyber} catalogued 24 major cybersecurity attacks and events on upstream systems. We have used this information as a baseline to present a timeline of chronological security incidents that have affected the upstream O\&G sector (see Figure \ref{fig:timeline_upstream_cyberattacks}). The temporal characteristic shows a growing frequency of data exfiltration attacks against upstream O\&G production companies in recent times which could be indicative of the consequences of increasing integration of real time OT monitoring parameters with corporate IT networks to improve decision making. This is a part of the digital oilfield trend being witnessed in the industry. The O\&G industry, in particular, is very competitive and almost any kind of leaked information can be beneficial to a competitor \cite{hacquebord2019drilling}. Obtaining sensitive data like well drilling techniques, data on suspected oil and gas reserves, and special recipes for premium products \cite{hacquebord2019drilling} including chemical injection and corrosion inhibitors can prove to be very valuable and therefore attractive to attackers.

\subsection{Threat Actors and Motivation}
Threat actors operating in this sector typically range from those looking for ransom to those operating for rivals within the industry or outside and last (but not the least) state sponsored agencies with specialised hackers at their disposal. The last category has immense resources and the potential to devastate critical infrastructure is massive \cite{ciepiela2017evolution} as seen in the case of Stuxnet - A virus that was reportedly designed by State intelligence to spy on and disrupt Iran's nuclear enrichment centrifuges, but also ended up spreading to infect Chevron facilities \cite{phys2012chevron}, a major O\&G company.

They can generally be classified as \cite{masood2016assessment} \cite{brode20217cyber} \cite{hacquebord2019drilling}:

\begin{itemize}
    \item Disgruntled Ex-Employee: Usually motivated by revenge on employer by triggering information disclosure to public to cause embarrassment, or to sell sensitive information. Person may still possess knowledge of sensitive information like passwords or system architecture.
    
    \item Insider Threat (Disgruntled Employee): Insider threat could also be motivated by revenge although there are several factors that could cause a person to turn against their employer. Defence against this kind of threat actor is very complex, as they have access to a lot of data.
    
    \item Hacktivists: This group is motivated by certain ideologies and will not hesitate to infiltrate a company they feel has gone against those principles. Their goals are usually to expose secrets and whistle-blowing.
    
    \item Nation State Hackers: These are hired by a Government to perform cyber operations against other nations. O\&G producing nations usually rely on the revenue generated from oil production as a major source of economic power. This is what makes the impact of successful attacks to be significant to victim States. These groups are highly resourceful and aim to inflict maximum damage (loss of life and damage to environment).
    
    \item Cyber Terrorists/Organised Crime: Non-State hackers are groups or individuals with the main intention of obtaining money by stealing sensitive data or confidential information and either selling it or blackmailing the company into paying a ransom.
    
\end{itemize}

There are also some known adversaries that have been identified to be targeting the O\&G sector. These include \cite{nygaard2020dragonstone}:

\begin{itemize}
    
    \item XENOTIME: This group has been known to target O\&G companies in the United States and Europe since 2018 and have compromised several ICS vendors and manufacturers.
    
    \item MANELLIUM: Since 2013, this group has been targeting petrochemical companies.
    
    \item CHRYSENE: Involved in the 2012 Shamoon cyberattack at Saudi Aramco and remains active and evolving in more areas.
    
    \item HEXANE: Capabilities of this group is still being studied by Dragos but was first identified in 2019.
    
    \item DYMALLOY: A highly aggressive and capable activity group that has the ability to achieve long term and persistent access to IT and OT for intelligence collection and possible future disruption events.
    
    \item APT33: A group that has compromised oil companies in the United States, Europe, and Asia by obscuring a dozen live C\&C (Command \& Control) servers that have been used to do reconnaissance and botnet management since 2018 \cite{hacquebord2019drilling}. C\&C connections to cloud services are difficulet to detect since they use normal services that any employee could use for legitimate purposes \cite{hacquebord2019drilling}.
    
\end{itemize}

In the next section, we will examine the unique challenges in the upstream O\&G sub-sector that has made it attractive for these threat actors to actively carry out attacks on the targets discussed.

\section{Challenges in securing upstream assets}
There are some unique traits that make the upstream O\&G sector more challenging to secure \cite{nygaard2020dragonstone} \cite{hacquebord2019drilling} when compared to other critical infrastructure industries. These are highlighted as:

\begin{itemize}
    \item Upstream assets are usually spread over huge geographical landscape, including significant assets offshore.
    \item Offshore assets are usually in remote locations and in deep waters.
    \item A large percentage of production facilities have been designed decades ago and lack modern security features which make them vulnerable and obvious targets for cyber attacks.
    \item The frequent integration of vendor systems with operating company systems.
    \item Dependencies: Large distances and deep waters make it costly to establish a computer network for offshore platforms. Frequent damage to fibre-optic cables on seabed make it challenging to establish redundant and completely independent network solutions.
\end{itemize}

Accuracy is also a big challenge in oil and gas as the exact amount/volumes of what is produced is not easily measured \cite{polyakov2015sap}. Hydrocarbon volumes fluctuate depending on environmental temperature and pressure conditions and require complex conversion calculations of the observed volumes at each custody transfer point \cite{polyakov2015sap}. It is possible to spoof this data in a way that will make it difficult to investigate \cite{polyakov2015sap}. Micro fractional changes to any one of the sensor parameters used in calculating hydrocarbon volumes over time could lead to significant losses to either operating companies or oil producing States. The latter could be better described as economic sabotage.

Process states and plant configurations are always changing, sometimes due to optimisations, but mostly as a result of degradation. An example is pressure vessels that have corroded beyond minimum thickness and can no longer withstand the Maximum Allowable Working Pressure (MAWP). Rather than outright replacement, vessels can be derated to a lower MAWP. This changes plant configuration and set point limits. Also, as discussed earlier, the life of a field in production may last up to 50 years, yet the assets have a design life considerably less than that - usually 25 years \cite{stacey2008life}. Life extension projects are carried out on facilities to extend, upgrade, and further optimise operations. This is why after a major maintenance phase, it is not unusual to have a system operating in a manner slightly different from prior to the maintenance activities. These configuration changes need to be taken into account when designing an efficient cyber security mitigation strategy.

The challenges in securing ICPS from cyber attacks in the offshore O\&G industry can also be broadly grouped into operational, financial, and legislative.

\textbf{Operational Challenges:} Keeping OT running at all times is critical for any successful industrial plant: every second systems are offline can cost the operating company thousands of dollars and recovering from a single hour offline can take days \cite{rosner2017using}. It is not unusual to witness systems running without being patched for years because operations availability and system up time go above security within ICS environment \cite{almusaher2020feasible}. In restarting production wells after a shut-in, producers must also weigh the cost and mechanical difficulty of restoring those wells back to pre-curtailed volumes \cite{adams2020oilwells} as the transfer of fluids back to the wellbore after production restoration is not usually very efficient or complete \cite{walser2021productionrestarts}. This creates a high risk scenario where after a significant shutdown, depending on the age of the well, previous production level may never be attained again. The industry is intolerant of frequent shutdowns and as a result, there aren't many opportunities for security updates and patches which are necessary as most of the offshore platforms are legacy systems. There is a high number of old offshore platforms still producing today. In fact, nine of the world’s longest-standing fixed offshore platforms are located in the North Sea, while one is in the Gulf of Mexico, US \cite{offshore2019longeststanding}. One of the oldest of them is a platform called Ekofisk 2/4 B, operated by ConocoPhillips and located 2.3km north of the Ekofisk Complex in the North Sea and it has been operating since 1974! To keep these platforms running efficiently, the operators retrofit new technologies onto legacy systems. This is usually done without security considerations that would adequately protect these systems from cyber attacks.

As discussed earlier, drilling campaigns are undertaken throughout the life of a field. Whenever a service company is contracted to drill wells for an operating company, this requires the use of shared computer networks, resulting in production equipment being exposed to network-related vulnerabilities \cite{nygaard2020dragonstone}. The frequent integration of vendor systems with operating company systems is another risk factor that increases the attack surface of O\&G production platforms. There is a need to ensure all sub-contractors keep the same or a higher level of cyber-hygiene than the operating company. 

\textbf{Financial Challenges:} The offshore O\&G industry is a highly regulated and capital intensive industry \cite{wanasinghe2020internet}. For example, FPSOs (Floating, Production, Storage, and Offloading vessels), because they give operators the freedom and versatility to explore remote areas and extract at a significantly cheaper cost \cite{tham2019fpso}, have become very popular in exploring deep offshore. The cost of a typical FPSO could range from \$800 million USD (Exxon Mobil's Kizomba A \cite{oil201910reasons}) to \$3 billion USD (Total Nigeria's Egina FPSO \cite{offshore2013samsung}). The upstream life cycle discussed earlier shows that the company bears these huge costs for a number of years (during the exploration, appraisal, and development phases) before production begins, and thus are trying to recoup huge investments made as quickly as possible during the production phase. In turn, the production phase is the most vulnerable to cyber attacks, but the company's focus during this phase is on trying to break even, turning a healthy profit before the reservoir is depleted (limited period), and meeting their obligations to the host Government through payment of taxes and royalties. To achieve this, continuous operations with minimal shutdowns is usually prioritised over security concerns.

\textbf{Legislative Challenges:} Several governments all over the world have recognised the threat that cybersecurity poses the critical infrastructure industry. Even though it is usually the norm that each individual company bears direct responsibility over securing its digital systems, severe cyber attacks will have national implications as well \cite{friis2018cyber}. This means that governments and the relevant agencies have a role to play in detecting, preventing and responding to such attacks \cite{friis2018cyber}. A seamless transition between private sector companies to authorities will require a holistic threat picture, clear areas of responsibilities, and good procedures that are exercised regularly. This is hardly the case today \cite{friis2018cyber}. In the United States, for example, There are stricter cybersecurity regulations that govern power, chemical, and nuclear facilities, but no federal laws impose such standards on the O\&G industry \cite{nygaard2020dragonstone}. O\&G companies are not required to report cyber incidents, and as a result, the specifics are usually kept secret because companies tend to disclose information in exchange for anonymity \cite{nygaard2020dragonstone}. This ensures that lessons learned from cyberattacks in one company and security measures implemented in response to such attacks are not always passed on to other companies in the sector, creating a serious knowledge gap \cite{nygaard2020dragonstone}. Attacks are getting more sophisticated and government legislation is playing catch up. There needs to be a concerted effort to create a legislative framework that ensures a minimum requirement for companies to secure their critical infrastructure assets from cyber attacks.

We have highlighted the financial and legislative challenges here to give context to the bigger industry problem, however neither are within the scope of this paper, as we have only focused on the operational challenges so far. In the next section, we will look at some general mitigation strategies, and how current datasets available to OT security researchers are inadequate for the o\&G industry.

\section{Securing Upstream O\&G Assets - Current State}

\subsection{General Mitigation Strategies}
General cyber security safeguards such as restricted physical access, cryptography, patch management, separation of corporate and production systems (through Demilitarized Zones (DMZ), Firewalls and Access Control Lists (ACLs)), and activity logging are all applicable mitigation strategies, but need to be viewed in conjunction with typical SCADA systems characteristics \cite{nazir2017assessing}. Although very little has focused on O\&G assets, in the broader context there are some practical applications that can improve the cyber hygiene of upstream assets. Esfahani et al. \cite{esfahani2017lightweight} and Srinivas et al. \cite{srinivas2018anonymous} are both studies that proposed the use of lightweight authentication to ensure only authorised users gain access. In \cite{esfahani2017lightweight}, a Machine-to-Machine (M2M)  protocol based on hash and XOR operations was applied in two phases - (a) the registration phase, where each smart sensor registers itself to an authentication server with replication of pre-shared keys with the router, and (b) the authentication phase where mutual authentication is achieved between the sensor and the router \cite{sengupta2020comprehensive}. \cite{srinivas2018anonymous} was based on chaotic map for IIoT environments which allows access to designated IoT devices only to authorised users with the use of personal biometric, smart cards, and passwords.

Research on ensuring basic security or defending against dreadful attacks in IIoT is still in its infancy \cite{sengupta2020comprehensive} especially for the O\&G sector, however, in the next sub-section, we shall examine intrusion detection systems and the limitation of datasets available to expand security research in this area that is applicable to the sector.

\subsection{Intrusion Detection Systems (IDS)}
Early warning and detection of breaches are essential for being in a state of readiness \cite{ciepiela2017evolution}. A number of systems have been designed to build threat detection capability in various industries, but only a few specifically for the O\&G industry. Amongst these, Al-Issa et al. \cite{al2012protecting} suggested using Network Behavior Anomaly Detection (NBAD) and Network Data Leak Prevention (NDLP) Systems to help detect unusual behaviours in O\&G facility systems and networks and detect the leak of information between the industrial control systems and the enterprise network respectively. Aljubran et al. \cite{aljubran2018integrated} also proposed three useful tools that could, at best, provide a partial solution to the cybersecurity threat faced by remote offshore oil facilities. These are Safety Instrumented Systems (SIS), decision tree, and risk management.

The recent trend for the vast majority of research on Intrusion Detection Systems (IDS) is by application of machine learning techniques in developing an IDS for ICS \cite{bhamare2020cybersecurity}.
Lack of adequate datasets remains the biggest hindrance to security research in this area is the. Machine learning (ML) has been used for identification of anomalous behaviours in industrial and manufacturing systems \cite{olowononi2020resilient}. A ML-based firewall suggested by Haghighi et al. \cite{haghighi2020machine} towards securing ICS was focused on accuracy and achieving zero false-positives in developed classifiers. In another example, Anthi et al. \cite{anthi2021hardening} explored how adversarial attacks can be used to target supervised classifiers by presenting generated adversarial DoS samples to a trained model and understanding their classification behaviours on IoT devices.

Bhamare et al. extensively reviewed related works in the field of securing ICS/SCADA from cyber threats using machine learning which is summarised in Table \ref{tab:ics_machine_learning} \cite{bhamare2020cybersecurity}. The studies were however limited in the scope of application as most were from specific industry data sets or limited simulated models of ICS that are not applicable to the O\&G industry. A summary of this is shown in Table \ref{tab:ml_techniques_in_ics}.

\begin{center}
\begin{table}[h]
\begin{tabular}{ | l | c | p{8.6cm} |}
\hline
 Authors & Reference & ML Model and Implementation\\
 \hline \hline
 Wehenkel & \cite{wehenkel1997machine} & Decision tree induction, multilayer perceptron and nearest neighbour classifiers \\
 \hline
 Dua and Du & \cite{dua2016data} & Cybersecurity using ML and data mining in general \\
 \hline
 Cardenas et al. & \cite{cardenas2011attacks} & Attack categorisation, IDS \\
 \hline
 Zhang et al. & \cite{zhang2013scada} & Support Vector Machine (S2 OCSVM), IDS \\
 \hline
 Yasakethu and Jiang & \cite{yasakethu2013intrusion} & Artificial Neural Network, Support Vector Machine, Hidden Markov Model \\
 \hline
 Beaver et al. & \cite{beaver2013evaluation} & Anomaly detection in SCADA via comparison of various ML algorithms \\
 \hline
 Maglaras and Jiang & \cite{maglaras2014intrusion} & One class Support Vector Machine, IDS \\
 \hline
 Hink et al. & \cite{hink2014machine} & OneR, NNge (Nearest Neighbour-like algorithm), Random Forests, Naive Bayes, SVM, JRipper, Adaboost \\
 \hline
 Erez and Wool & \cite{erez2015control} & Single window classification algorithm deployed on IDS to detect irregular changes in SCADA control register values \\
 \hline
 Franc et al. & \cite{franc2015learning} & A Multiple Instance Learning algorithm used on network logs for security \\
 \hline
 Nader et al. & \cite{nader2016detection} & ML techniques with kernel methods to detect cyber attacks in water distribution systems \\
 \hline
 leahy et al. & \cite{leahy2016diagnosing} & Classification ML techniques \\
 \hline
 Valdes et al. & \cite{valdes2016anomaly} & Unsupervised ML methods for anomaly detection in electrical substation circuits \\
 \hline
 Stefanidis and Voyiatzis & \cite{stefanidis2016hmm} & Hidden Markov Model, IDS \\
 \hline
 Bartos et al. & \cite{bartos2016optimized} & Support Vector Machine-based classification system \\
 \hline
\end{tabular}
\caption{ICS/SCADA Cybersecurity: Summary of Machine Learning Approaches \cite{bhamare2020cybersecurity}}
\label{tab:ics_machine_learning}
\end{table}
\end{center}

While contemporary IDSs use machine learning algorithms for pattern recognition to detect threat activities that are anomalous for a particular system, there are other IDSs which use signature-based systems to compare the activities to a database of known threats \cite{yasakethu2013intrusion}, \cite{maglaras2014intrusion}, \cite{dua2016data}, \cite{bhamare2020cybersecurity}. Both of these methods can be combined to develop a robust detection system \cite{bhamare2020cybersecurity}.

Zeng et al. \cite{zeng2018intrusion} introduced a taxonomy of detection approach and also discussed machine learning-based solutions along with other types of available approaches for IDSs deployed in ICS \cite{zeng2018intrusion} \cite{bhamare2020cybersecurity}. By their own admission, the authors confirmed that from the papers they surveyed, power systems are the main field that investigators study in \cite{zeng2018intrusion}. From their comparison, most of the datasets or testbeds utilised were from power grids and a small percentage from water distribution systems.

\begin{table}[]
\resizebox{\textwidth}{!}{%
\begin{tabular}{@{}lll@{}}
\toprule
ML Technique & Authors & Domain Secured \\ \midrule
SVM/OCSVM & \cite{beaver2013evaluation} \cite{he2017real} \cite{potluri2017evaluation} \cite{keliris2016machine} \cite{yasakethu2013intrusion} \cite{zhang2010mitigating} \cite{maglaras2014intrusion} & integrity, availability, confidentiality \\
Naïve Bayes & \cite{ullah2017hybrid} \cite{beaver2013evaluation} & integrity, confidentiality \\
Decision Trees/Random Forests & \cite{beaver2013evaluation} \cite{ullah2017hybrid} \cite{siddavatam2017ensemble} & integrity, confidentiality \\
Deep Belief Network & \cite{he2017real} \cite{potluri2017evaluation} & availability, integrity \\
Artificial Neural Network & \cite{he2017real} \cite{yasakethu2013intrusion} & integrity \\
KNN/K-means & \cite{alves2018embedding} \cite{eigner2016detection} & authentication, confidentiality, availability, integrity \\ \bottomrule
\end{tabular}%
}
\caption{Popular ML techniques used in ICS Security}
\label{tab:ml_techniques_in_ics}
\end{table}

Challenges in the way of utilising machine learning and how it can help in defence mechanisms with respect to the relevant threats in ICS have been reviewed comprehensively by Zolanvari et al. \cite{zolanvari2019machine}, \cite{bhamare2020cybersecurity}. A case study was also presented where a ML-based IDS was developed using a SCADA testbed. The dataset from the testbed was deliberately built to be imbalanced by making the percentage of attack traffic in the dataset less than 0.2\%.

A comparative analysis of various ICS datasets, summarised in Table \ref{tab:ics_datasets}, was carried out by Choi et al. \cite{choi2018comparison}. The analysis seems to agree with our observed limitations of the current datasets used to conduct ICS security research and highlights why most are not applicable to a broad set of scenarios. For our case specifically (O\&G offshore industry), most of the datasets do not account for the dynamic behaviour of monitored variables identified earlier. Pressure and temperature values change throughout the life of a producing field as the reservoir is being depleted which result in different hydrocarbon volumes calculated at any point in time. The monitored variables in current datasets only fluctuate within a given range. This is summarised in Table \ref{tab:ICS_data_capture_summary}.

The review of existing literature shows that although a lot of research has been conducted on IDS security, the common limitation has been the availability of a wide scope dataset that applies to several critical infrastructure industries. The power industry is the most represented sector while an opportunity exists to create new datasets that represents commonly deployed ICS setup in the O\&G industry.

\begin{center}
\begin{table}[h]
\begin{tabular}{ | l | p{2cm} | p{2.4cm} | p{1.4cm} | c | p{1.3cm} |}
\hline
 ICS Dataset & Protocols & System & Year of release & Data-type & Reference\\
 \hline \hline
 Morris et al. & Modbus & Power, water, gas & 2013, 2014, 2015, 2017 & csv, arff & \cite{morris2014industrial} \\
 \hline
 Lemay & Modbus & SCADA sandbox & 2016 & csv, pcap & \cite{lemay2016providing} \\
 \hline
 SWaT & Modbus, Ethernet/IP & Water treatment & 2016 & csv & \cite{goh2016dataset} \\
 \hline
 Rodofile et al. & S7Comm & Mining refinery & 2017 & csv, pcap & \cite{rodofile2017process} \\
 \hline
 4SICS & Modbus, S7Comms, DNP3, Ethernet/IP & Complex & 2015 & pcap & \cite{icslab20154sics} \\
 \hline
 S4x15 ICS Village CTF & Modbus & Complex & 2015 & pcap & \cite{peterson2015s4x15} \\
 \hline
 DEFCON 23 ICS Village & Modbus & Complex & 2015 & pcap & \cite{defcon232015} \\
 \hline
\end{tabular}
\caption{Summary of ICS datasets publicly available \cite{choi2018comparison}}
\label{tab:ics_datasets}
\end{table}
\end{center}

\begin{table}[]
\begin{tabular}{|p{1.3cm}|l|p{1.3cm}|l|llp{2cm}|}
\hline
\multicolumn{1}{|c|}{\multirow{2}{*}{\textbf{ICS   Dataset}}} &
  \multicolumn{1}{c|}{\multirow{2}{*}{\textbf{Num. of Pkts}}} &
  \multicolumn{1}{c|}{\multirow{2}{*}{\textbf{Byte of Pkts}}} &
  \multicolumn{1}{c|}{\multirow{2}{*}{\textbf{Duration}}} &
  \multicolumn{3}{c|}{\textbf{Data Capture}} \\ \cline{5-7} 
\multicolumn{1}{|c|}{} &
  \multicolumn{1}{c|}{} &
  \multicolumn{1}{c|}{} &
  \multicolumn{1}{c|}{} &
  \multicolumn{1}{l|}{\textbf{Continuous}} &
  \multicolumn{1}{l|}{\textbf{Interruptions}} &
  \textbf{Dynamic Variables} \\ \hline
Lemay             & 2,588,491  & 169,690,458   & 15 hours    & \multicolumn{1}{l|}{No}  & \multicolumn{1}{l|}{Yes} & No \\ \hline
SWaT              & 19,761,714 & 5,498,545,489 & 11days      & \multicolumn{1}{l|}{Yes} & \multicolumn{1}{l|}{No}  & No \\ \hline
Rodofile          & 23,387,064 & 5,848,801,728 & 27 hours    & \multicolumn{1}{l|}{Yes} & \multicolumn{1}{l|}{No}  & No \\ \hline
4SICS             & 3,773,984  & 314,562,089   & 1d 22 h 7 m & \multicolumn{1}{l|}{Yes} & \multicolumn{1}{l|}{No}  & No \\ \hline
S4x15CTF DEFCON23 & 1,678,668  & 124,271,095   & N/A         & \multicolumn{1}{l|}{Yes} & \multicolumn{1}{l|}{No}  & No \\ \hline
\end{tabular}
\caption{ICS datasets: Data capture summary}
\label{tab:ICS_data_capture_summary}
\end{table}


\subsection{Overcoming the Challenge of Insufficient Research Data}
Typically, industrial control systems manage critical industries (oil and gas, water, power, and chemical industries) that are sensitive to downtime. The risk of carrying out penetration tests on live operational equipment in the field to validate intrusion detection systems or other measures to improve the cybersecurity of such facilities is too high and can potentially cause damage or lead to loss of lives. Testbeds are the preferred methods for emulating ICS and deploying cyber threat detection/protection models on. However, the working philosophy and behaviour of different industry verticals is different. To design and develop the industry-specific best fit solution requires testbed of each type \cite{stojanovic2020challenges}. This will help in creating varied datasets that include industry-specific scenarios combined with machine learning models that are trained on benign and malicious activities which could improve the security posture of the industry.

\section{Conclusion}
This paper reviewed the growing threat of cyber attacks to ICPS in the offshore O\&G industry as a result of the advancements in technology, digitalisation and integration of oil field equipment with corporate networks, and the need for remote monitoring and control. This has increased the attack surface available for cyber attackers to exploit. A timeline of documented cyber attacks on upstream O\&G assets was presented which showed that data exfiltration has become more common in recent times, coinciding with the increase in integration of OT equipment with IT networks that is now prevalent in the industry. Furthermore, we gave a brief description of offshore O\&G operations and the oil and gas production process, highlighting the possible areas of cyber attack infiltration.

We analysed a typical subsea control system architecture and highlighted its vulnerabilities to MiTM, DoS, and spoofing attacks by mapping the attacks to one or more layers of the architecture. Correlating these to documented cyber security incidents that affected the upstream O\&G industry in recent times showed that threat actors have the capability to breach subsea control systems in its current state. We also discussed challenges in securing upstream assets, highlighting dynamic process state changes due to operations like derating of pressure vessels and asset life extension projects which add to the complexity of identifying whether a changed plant configuration is legitimate or due to malicious actors. Mitigating strategies were also highlighted involving the use of IDS. There remains a lack of adequate datasets representative of processes in upstream oil and gas production.

\subsection{Future Directions}
Looking at the shift in philosophy of offshore installations from manned to unmanned facilities to be controlled remotely, and the current research into designing several of such, we can safely deduce that there will be an increase in digitalisation of offshore oilfields. With ICPS and IIoT improving integration of OT and IT there will be lots of opportunities for studies discovering new vulnerabilities in new systems. However, there is a need to create new datasets that also cater for the dynamic measurements and sensor readings of the offshore O\&G industry. This shows that ICPS security in the O\&G sector will continue to be a well researched area for the foreseeable future.

\bibliographystyle{unsrt}  
\bibliography{references}  






\end{document}